\documentstyle[psfig,aps]{revtex}

\catcode`\@=11
\def\marginnote#1{}

\def\ifmath#1{\relax\ifmmode #1\else $#1$\fi}

\def\bold#1{\setbox0=\hbox{$#1$}%
     \kern-.025em\copy0\kern-\wd0
     \kern.05em\copy0\kern-\wd0
     \kern-.025em\raise.0433em\box0 } 

\def\GENITEM#1;#2{\par\vskip6pt \hangafter=0 \hangindent=#1
   \Textindent{$ #2$ }\ignorespaces}

\newcount\hour
\newcount\minute
\newtoks\amorpm
\hour=\time\divide\hour by60
\minute=\time{\multiply\hour by60 \global\advance\minute by-
\hour}
\edef\standardtime{{\ifnum\hour<12 \global\amorpm={am}%
    \else\global\amorpm={pm}\advance\hour by-12 \fi
    \ifnum\hour=0 \hour=12 \fi
    \number\hour:\ifnum\minute<100\fi\number\minute\the\amorpm}}
\edef\militarytime{\number\hour:\ifnum\minute<100\fi\number\minute}
\def\draftlabel#1{{\@bsphack\if@filesw {\let\thepage\relax
  \xdef\@gtempa{\write\@auxout{\string
    \newlabel{#1}{{\@currentlabel}{\thepage}}}}}\@gtempa
    \if@nobreak \ifvmode\nobreak\fi\fi\fi\@esphack}
     \gdef\@eqnlabel{#1}}
\def\@eqnlabel{}
\def\@vacuum{}
\def\draftmarginnote#1{\marginpar{\raggedright\scriptsize\tt#1}}
\def\draft{\oddsidemargin -.5truein
        \def\@oddfoot{\sl preliminary draft \hfil
        \rm\thepage\hfil\sl\today\quad\militarytime}
        \let\@evenfoot\@oddfoot \overfullrule 3pt
        \let\label=\draftlabel
        \let\marginnote=\draftmarginnote

\def\@eqnnum{(\theequation)\rlap{\kern\marginparsep\tt\@eqnlabel}%
\global\let\@eqnlabel\@vacuum}  }
\def\preprint{\twocolumn\sloppy\flushbottom\parindent 1em
        \leftmargini 2em\leftmarginv .5em\leftmarginvi .5em
        \oddsidemargin -.5in    \evensidemargin -.5in
        \let\@evenfoot\@oddfoot \overfullrule 3pt
        \let\label=\draftlabel
        \let\marginnote=\draftmarginnote

\def\@eqnnum{(\theequation)\rlap{\kern\marginparsep\tt\@eqnlabel}%
\global\let\@eqnlabel\@vacuum}  }
\def\preprint{\twocolumn\sloppy\flushbottom\parindent 1em
        \leftmargini 2em\leftmarginv .5em\leftmarginvi .5em
        \oddsidemargin -.5in    \evensidemargin -.5in
        \columnsep 15mm \footheight 0pt
        \textwidth 250mmin      \topmargin  -.4in
        \headheight 12pt \topskip .4in
        \textheight 175mm
        \footskip 0pt

\def\@oddhead{\thepage\hfil\addtocounter{page}{1}\thepage}
        \let\@evenhead\@oddhead \def\@oddfoot{} \def\@evenfoot{}
}
\def\titlepage{\@restonecolfalse\if@twocolumn\@restonecoltrue\onecolumn
     \else \newpage \fi \thispagestyle{empty}\c@page\z@
        \def\thefootnote{\fnsymbol{footnote}} }
\def\endtitlepage{\if@restonecol\twocolumn \else  \fi
        \def\thefootnote{\arabic{footnote}}
        \setcounter{footnote}{0}}  
\catcode`@=12
\relax

\def\be{\begin{equation}}
\def\ee{\end{equation}}
\def\br{\begin{eqnarray}}
\def\er{\end{eqnarray}}

\def\NPB#1#2#3{{\it Nucl.~Phys.} {\bf{B#1}} (19#2) #3}
\def\PLB#1#2#3{{\it Phys.~Lett.} {\bf{B#1}} (19#2) #3}
\def\PRD#1#2#3{{\it Phys.~Rev.} {\bf{D#1}} (19#2) #3}
\def\PRL#1#2#3{{\it Phys.~Rev.~Lett.} {\bf{#1}} (19#2) #3}

\def\PRD#1#2#3{{\it Phys.~Rev.} {\bf{D#1}} (19#2) #3}
\def\PRL#1#2#3{{\it Phys.~Rev.~Lett.} {\bf{#1}} (19#2) #3}

\begin{document}

\topmargin-1.5cm  

\begin{titlepage}
\begin{flushright}
SUSX-TH/98-011\\ 
IEM-FT-174/98\\
\end{flushright}
\vspace{.2in}
\begin{center}

{\large{\bf Stabilizing the Dilaton in Superstring Cosmology}
}
\bigskip \\
{\large T.~Barreiro\footnote{Email address: 
mppg6@pcss.maps.susx.ac.uk}, 
B.~de~Carlos\footnote{Email address: B.de-Carlos@sussex.ac.uk} and 
E.~J.~Copeland\footnote{Email address: E.J.Copeland@sussex.ac.uk}\\ 
\vskip 0.2in
{\it 
Centre for Theoretical Physics, University of Sussex, \\ Falmer, 
Brighton BN1 9QH, United Kingdom \\}
}
\vskip 0.2in
{\bf Abstract} \smallskip \end{center} \setcounter{page}{0}
We address the important issue of stabilizing the dilaton in the 
context of superstring cosmology. Scalar potentials which arise out of 
gaugino condensates in string models are generally exponential in 
nature. In a cosmological setting this allows for the existence of 
quasi scaling solutions, in which the energy density of the scalar 
field can, for a period, become a fixed fraction of the background 
density, due to the friction of the background expansion. Eventually 
the field can be trapped in the minimum of its potential as it 
leaves the scaling regime. We investigate this possibility in various 
gaugino condensation models and show that stable solutions for the 
dilaton are far more common than one would have naively thought.

\end{titlepage}

\newpage
 
\section{Introduction}
Scalar fields in cosmology have been extensively studied over the past
few years. One of the most intriguing areas in which they occur is 
Superstring theory, where the presence of the dilaton field is vital. 
Its vacuum expectation value (VEV) determines both the gauge and 
gravitational coupling constants of the low energy theory and also 
fixes the scale of supersymmetry (SUSY) breaking through the gravitino
mass, $m_{3/2}$. Therefore realistic models require a VEV of order one
(in Planck units), and $m_{3/2} \sim 1$ TeV.   

Unfortunately, in string theory the dilaton potential is flat to all 
orders in perturbation theory, which of course means there is no way
of obtaining a stable VEV for the field. This problem has to be 
overcome through some nonperturbative effect. The most promising 
possibility is through the formation of condensates of gaugino 
fields at an energy scale of around $10^{14}$ GeV \cite{deren85}. 
The resulting scalar potential for the dilaton is then a combination 
of exponentials and polynomials in the field. A detailed investigation
of these condensate models has demonstrated the need for at least two 
condensates to form if the dilaton potential is to develop a minimum at
a realistic value although with a negative cosmological constant
(these are the so-called ``racetrack'' models) \cite{decar93}.

An alternative proposal has recently been suggested as a method of 
obtaining a minimum for the dilaton field, and it has the advantage of 
relying on only one gaugino condensate \cite{casas96,binet97}. In this
scenario the K\"ahler potential (which determines the kinetic terms of
the dilaton in the action) requires string inspired nonperturbative 
corrections. A detailed analysis of these models \cite{barre97}
indicates that it is possible to have a minimum with zero or small 
positive cosmological constant. One additional positive feature that 
emerges is that the nonperturbative corrections can lead to a
solution of the ``moduli problem'' for the dilaton \cite{decar93p}
(fields with masses in the TeV range but which decay so slowly that 
they spoil nucleosynthesis), by giving it a huge mass.

Although attractive, both kind of models still have several problems 
associated with them. One is the difficulty of achieving 
inflation which  was carefully studied by Brustein and Steinhardt 
\cite{brust93} a few years ago. Taking a model of multiple gaugino 
condensation (as those studied in \cite{decar93}) they argued that the
kinetic energy associated  with the dilaton field would dominate over 
its potential energy until $\phi$ (the canonically normalized field, 
related to the usual dilaton by ${\rm Re} S = e^{\phi}$) would settle 
near a minimum of its potential. This obviously excludes inflation 
from happening, at least with the dilaton as the inflaton field, 
leaving the possibility of $\phi$ settling down to a minimum and then 
inflation being driven by other fields. However this second 
possibility also presented serious problems as the models they studied
had a negative cosmological constant. Nonperturbative corrections to 
the K\"ahler potential can cure this latter problem, but in both cases
the potentials are exponentially steep in the strong coupling regime. 
This would lead us to expect the dilaton to roll past the minimum 
rather than acquiring its VEV, which seems to be a major problem that 
superstring cosmology needs to address.

In this paper we turn our attention to the possibility that other 
matter fields rather than the dilaton drive the evolution of the 
Universe. Recent attention in cosmology has turned to the 
investigation of scaling solutions in models with exponential scalar 
field potentials \cite{wette88,ferre97,copel98}. These models are of 
particular interest because if the background dynamics are dominated 
by some matter source other than the field itself (i.e. radiation, 
dust, vacuum energy) then it is possible for the field to enter a 
scaling regime as it evolves down its potential. In this regime the 
friction term from the expansion of the Universe balances the kinetic 
energy of the field allowing it to enter this scaling era. Attractor 
solutions exist \cite{wette88,copel98} where the energy density in the
field becomes a fixed fraction of the total energy density. 

This intriguing behaviour can be applied to the case of the dilaton 
field arising from string theory. Under the assumption that it is 
evolving from somewhere in the strong coupling regime of its potential
we show how, for a wide range of initial field values in the presence 
of a background dominated by a barotropic fluid, the dilaton enters a 
quasi scaling regime as it evolves down the potential, inspite of its 
steepness. This scaling behaviour eventually ends as the field enters 
the minimum of its potential, by which time it has slowed down 
sufficiently for it to simply oscillate about it, losing energy and 
eventually becoming fixed with a realistic VEV. 

In section II we introduce the concept of scaling solutions with 
exponential potentials. In section III we demonstrate how this can be 
successfully adapted to both the racetrack and modified K\"ahler 
potential models of gaugino condensates. Solutions are presented 
analytically and numerically showing how the dilaton field is 
stabilized in its minimum. We conclude in section IV.      

\section{Scaling solutions with exponential potentials}

In this section we will describe some general features concerning the 
cosmological evolution of scalar fields with exponential potentials.
Let us consider a scalar field $\phi$ with a potential energy density 
given by $V = V_0 e^{-\lambda \kappa \phi}$, with $\kappa^2 \equiv
8\pi G$ and $\lambda$ and $V_0$ constants, which is evolving in a 
Friedmann-Robertson-Walker (FRW) Universe containing a fluid with 
barotropic equation of state $p_{\gamma} = (\gamma-1) \rho_{\gamma}$, 
where $\gamma$ is a constant ($0 \leq \gamma \leq 2$, for instance 
$\gamma=4/3$ for radiation domination or $\gamma=1$ for matter 
domination). The equations of motion for a spatially flat FRW model 
with Hubble parameter $H$ are
\br
\dot{H} & = & -\frac{\kappa^2}{2} (\rho_{\gamma} + p_{\gamma} + 
\dot{\phi^2}) \nonumber \\
\dot{\rho}_{\gamma} & = & -3 H (\rho_{\gamma} + p_{\gamma}) 
\label{system}\\
\ddot{\phi} & = & -3 H \dot{\phi} - \frac{dV}{d \phi} \;\;, 
\nonumber
\er
subject to the constraint
\be
H^2 = \frac{\kappa^2}{3} (\rho_{\gamma} + \frac{1}{2} \dot{\phi^2} 
+ V) \;\;.
\label{const}
\ee
In the previous equations we have assumed that the only interaction 
between $\phi$ and the other matter fields is gravitational. In what 
follows we set $\kappa^2 =1$ but it can easily be reinstated. For 
exponential potentials the asymptotic behaviour of this system can be 
obtained analytically, and we shall focus on the solution for very 
steep potentials, namely $\lambda^2 > 3\gamma$, as these are of most 
interest in string theory. Let us first of all review the structure of
the solutions which, as is well known \cite{wette88}, contain a 
late-time attractor solution. For that purpose it is useful to proceed
as in \cite{copel98}, defining the variables 
$x \equiv {\dot{\phi}}/{\sqrt{6}H}$ and $y \equiv 
{\sqrt{V}}/{\sqrt{3}H}$ and using the logarithm of the scale factor 
$N \equiv \ln(a)$ as the time variable. The previous system 
Eq.~(\ref{system}) becomes
\br
x' & = & -3x + \lambda \sqrt{\frac{3}{2}} y^2 + \frac{3}{2} x [2x^2+
\gamma (1-x^2-y^2) ] \nonumber \\
y' & = & -\lambda \sqrt{\frac{3}{2}} xy +  \frac{3}{2} y [2x^2+
\gamma (1-x^2-y^2) ] \label{systemr} \\
H' & = & - \frac{3}{2} H [2x^2+ \gamma (1-x^2-y^2) ] \;\;, \nonumber
\er
where a prime denotes a derivative with respect to $N$. In terms of 
these variables, the constraint Eq.~(\ref{const}) becomes 
$x^2 + y^2 + {\rho_{\gamma}}/{3H^2}=1$ or, in other words, for
$\rho_{\gamma} \geq 0$ we have the bounds $0 \leq x^2+y^2 \leq 1$.

\begin{figure}
\centerline{  
\psfig{figure=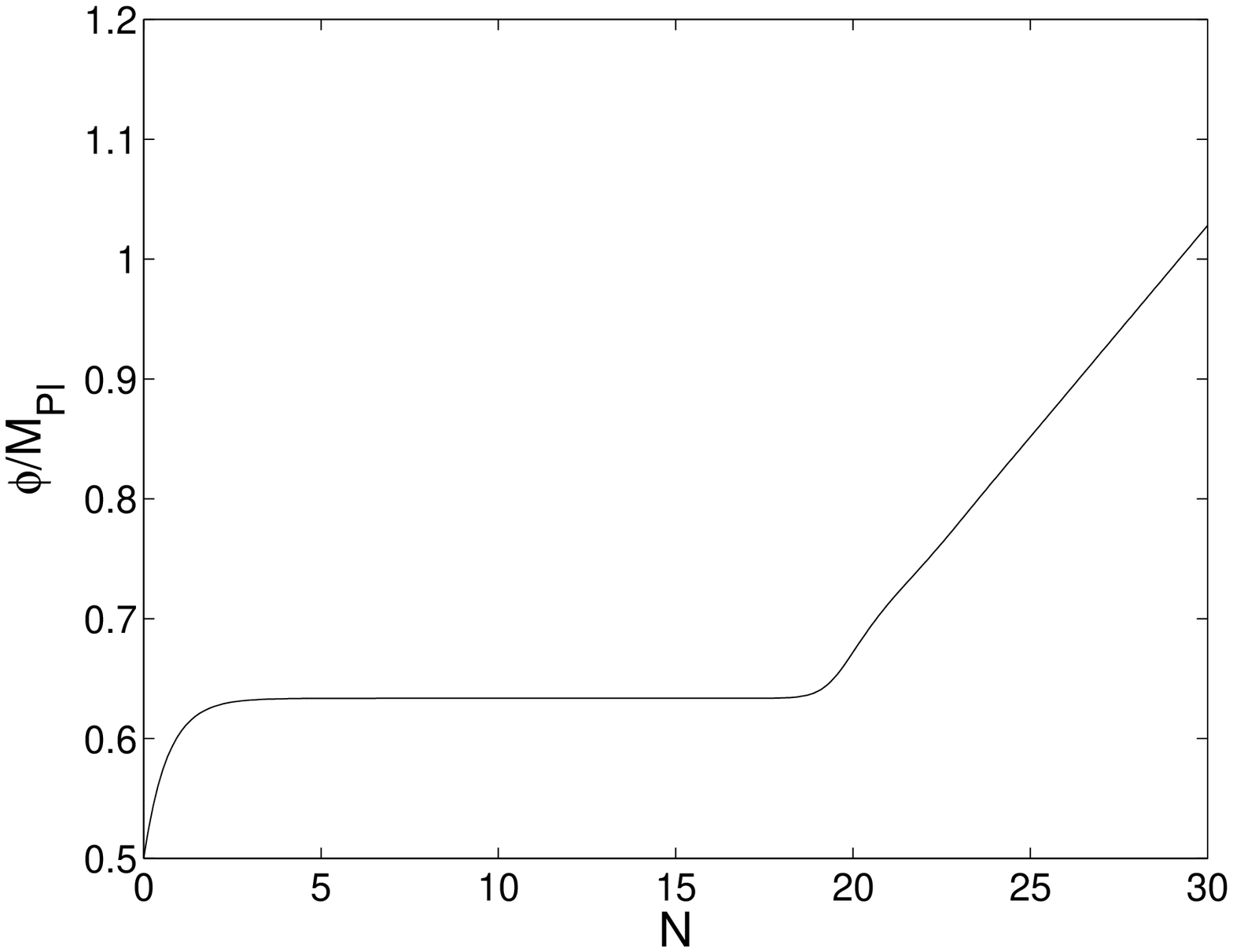,height=9cm,width=9cm,bbllx=0cm,bblly=7cm,bburx=21cm
,bbury=21cm}
}
\caption{}
{\footnotesize Plot of the evolution of $\phi$ versus $N=\ln(a)$. The
scalar potential is given by $V=e^{-\lambda \phi}$ with $\lambda=85$,
$\gamma=1$
and the initial conditions are $\phi_0=0.5$, 
$\dot{\phi_0}=0.2$ and $H_0=1$.
}
\end{figure}

The evolution of the $\phi$ field in terms of $N$ can be seen in 
Fig.~1, for $\lambda=85$, $V_0=1$ and initial conditions 
$\phi_0=0.5$, $\dot{\phi_0}=0.2$, and $H_0=1$. After an initial 
increase driven by the initital velocity of the field, we see how 
the friction term (or, in other words, the expansion of the 
Universe) dominates and freezes $\phi$ at a constant value for a 
considerable amount of time, until the field reaches a scaling regime 
corresponding to the critical points $x_c = \sqrt{3/2} \gamma/\lambda$ 
and $y_c = \sqrt{3(2-\gamma) \gamma/2\lambda^2}$ \cite{copel98}. 
This behaviour can also be obtained by solving the system of equations
Eq.~(\ref{systemr}) under the following assumptions:

\begin{itemize}

\item Stage~I (pre scaling regime): Since $\lambda \gg 1$, and $H$ is 
decreasing, $y$ is very small in this early stage of the evolution and
can be neglected in Eq.~(\ref{systemr}). The equation for $x$ now 
becomes
\be 
x' = -3x + \frac{3}{2} x [2x^2+\gamma(1-x^2)] \;\;,
\ee
and its solution is
\be
x = \left( 1+\frac{1-x_0^2}{x_0^2} e^{3(2-\gamma)N} \right)^{-1/2} \;\;, 
\label{xsol}
\ee
where $x_0$ is the initial condition for $x$ (at $N=0$). The solution 
for $\phi$ can be obtained by integrating Eq.~(\ref{xsol}) (recall
that $x = \phi'/\sqrt{6}$), and is given by
\be
\phi_{\rm I} (N) = \phi_{0} + \frac{2 \sqrt{6}}{3(2-\gamma)} 
\left[ \sinh^{-1} \left( \frac{x_0}{\sqrt{1-x_0^2}} \right) 
- \sinh^{-1} \left( \frac{x_0}{\sqrt{1-x_0^2}} e^{-3(2-\gamma)N/2} 
\right) \right] \;\;,
\label{stageI}
\ee
where $\phi_0$ is the initial value of the field. As $N$ increases
this solution tends to a constant value $\tilde{\phi_0}$ given by 
\be  
\tilde{\phi_0} = \phi_0 + 
\frac{\sqrt{6}}{3(2-\gamma)}
\ln \left( {\frac{1+x_0}{1-x_0}} \right) \;\;. \label{phitil}
\ee
Obviously for zero initial velocity, $\tilde{\phi_0}$ will reduce to 
$\phi_0$.

\item Stage~II (scaling regime): as mentioned above, the scaling
regime is defined by constant (i.e. critical) values for $x$ 
and $y$. We then obtain $H=H_0 e^{-3\gamma N/2}$ and
\be
\phi_{{\rm II}}(N) = \frac{1}{\lambda} \ln \left( 
\frac{2 \lambda^2 V_0}{9H_0^2 (2-\gamma) \gamma}\right) + 
\frac{3\gamma}{\lambda} N \label{scaling} \;\;,
\ee
where now we have an explicit dependence on the characteristics of 
the potential (i.e., $\lambda$ and $V_0$).

\end{itemize}

It is important to note that the background evolution is being 
determined by the additional matter fields present, given by
$\gamma$. It is not the $\phi$ field which is driving the evolution.
The fact that the simultaneous evolution of the scalar field together
with the background causes the former to reach a scaling regime,
inspite of the steepness of its (exponential) potential, suggests that
this could be also the case for the kind of potentials arising from
SUSY breaking via gaugino condensation, where the superpotential
depends exponentially on the dilaton field. In the next section we
will apply these solutions to the gaugino condensation models studied 
in \cite{decar93} and \cite{barre97}.

\section{Scaling solutions with gaugino condensates}
\subsection{Two condensate potentials}

Multiple gaugino condensation (or racetrack) models have been
extensively studied in the literature \cite{krasni87,decar93}. 
Essentially the idea is to consider a strong type interaction in the 
hidden sector of our effective supergravity (SUGRA) theory, which is 
governed by a nonsemisimple gauge group. The superpotential of such 
models will then be expressed in terms of a sum of exponentials which 
conspire to generate a local minimum for the dilaton. To be more
precise, the scalar potential in any N=1 SUGRA model \cite{crem83} is 
given by
\be
V = e^K |W|^2 \left[ \left(K^i + \frac{W^i}{W} \right) (K_i^j)^{-1}
\left( K_j + \frac{\bar{W}_j}{\bar{W}} \right) - 3 \right] \;\;,
\label{pot}
\ee
where $K$ is the K\"ahler potential, $W$ is the superpotential and the
subindices $i$, $j$ represent derivatives of these two functions with 
respect to the different fields. Given that we are interested in
superstring derived models, and in particular in studying the hidden
sector of the theory, both $K$ and $W$ will be dependent on the
dilaton ($S$) and the moduli ($T_i$, $i=1,2,3$) fields. In fact we
know that in the case of orbifold compactifications the tree-level 
K\"ahler potential is given by:
\be
K = -\log(S+\bar{S}) - \sum_{i=1}^{3} \log(T_i+\bar{T_i}) \;\;,
\label{kahler}
\ee
and we will restrict our study to the case of a hidden sector
interaction governed by two gauge groups, SU($N_1$) $\times$ 
SU($N_2$) under which we have $M_1(N_1+\bar{N}_1)$ and 
$M_2(N_2+\bar{N}_2)$ ``quark'' representations with Yukawa couplings 
to a set of singlet fields. For simplicity, we will assume an overall 
modulus $T = T_1 = T_2 = T_3$ and a generic singlet field for each of 
the gauge groups, $A_1$ and $A_2$ respectively. In this case, the 
superpotential is given by
\be
W = \sum_{i=1}^{2} \left[ -\frac{d_i} {\eta(T)^{\beta_i}}
 A_i^{M_i/N_i} e^{- \alpha_i S}    
+ h_i A_i^3 \right] \;\;, 
\label{sup} 
\ee
where $\alpha_i = 8 \pi^2/N_i$, $\beta_i= 2 (3 N_i - M_i) / N_i$, 
$d_i = N_i (32 \pi^2 e)^{(M_i/N_i -1)}$, $\eta(T)$ is the Dedekind 
function and the self coupling of the $A_i$ field is set to $h_i=1$.
[Note $N_i$ is not the same as $N$, the number of e-foldings defined 
earlier].

These kind of models were thoroughly studied in \cite{decar93},
so let us summarize their main features: the presence of the $\eta$
function, imposed by the requirement of target space modular 
symmetry and, in general, the $T$-dependence of the potential 
ensures the presence of a minimum for $T\sim 1.2$ (in Planck units)
\cite{font90}, independently of the particular gauge groups and/or 
matter representations (provided the dilaton acquires a VEV); also 
there exist minima in the ${\rm Im} S$ direction
if both condensates have opposite phases. Finally it was shown the
existence of many examples for which there is a minimum in the 
${\rm Re}S$ direction at the phenomenologically acceptable value 
${\rm Re} S \sim 2$ (remember that ${\rm Re S} = g^{-2}_{string}$), 
with a reasonable ($\sim 1$ TeV) gravitino mass but always with a 
negative value of the potential energy. A typical example is shown 
in Fig.~2(a), which perfectly illustrates the problem that Brustein 
and Steinhardt pointed out in their paper: the steepness of the 
dilaton potential, which would prevent the field from settling down 
at its (negative) minimum, instead allowing it to run over the tiny 
maximum towards infinity. 

\begin{figure}
\centerline{
\psfig{figure=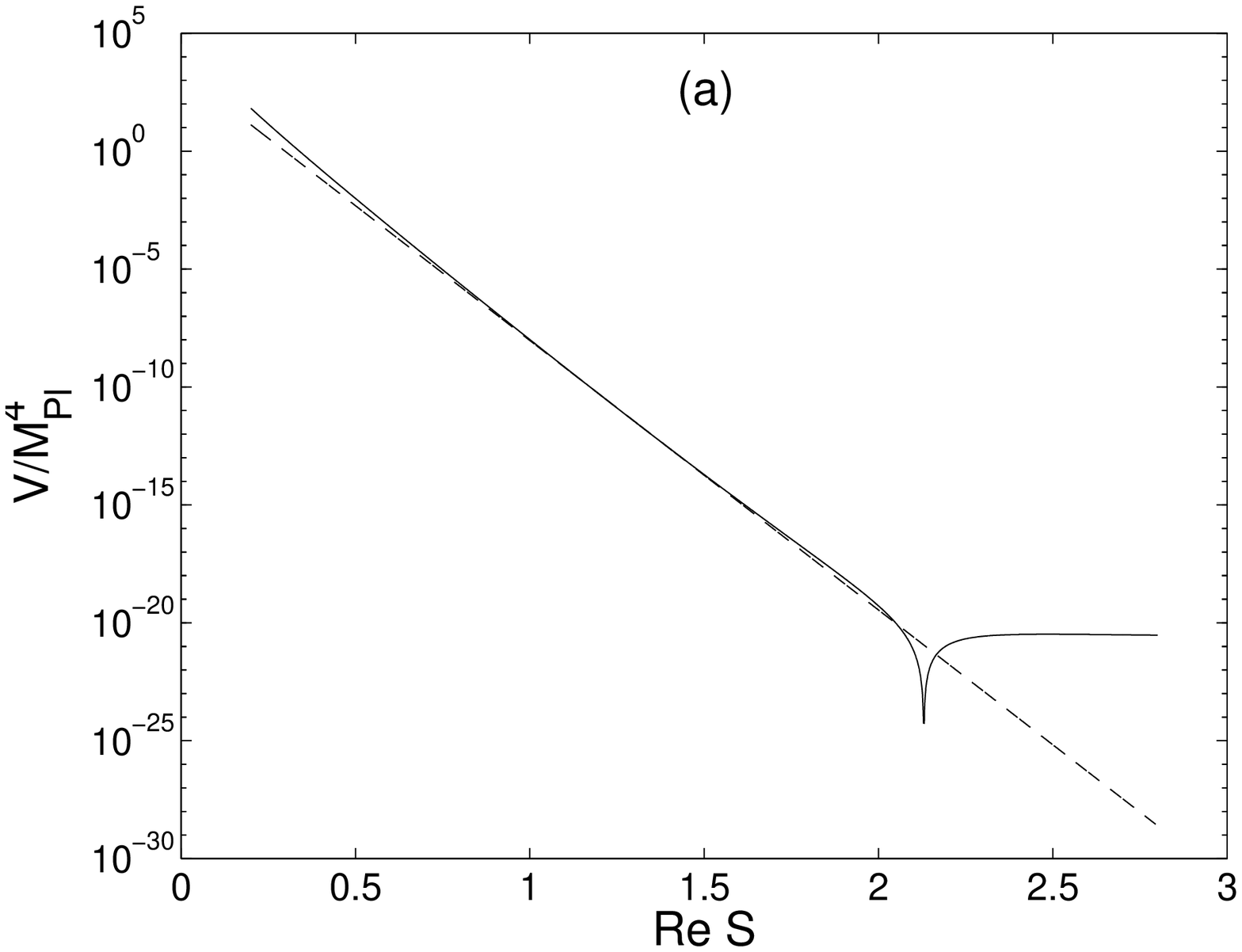,height=9cm,width=9cm,bbllx=0cm,bblly=7cm,bburx=21cm
,bbury=21cm}\
\psfig{figure=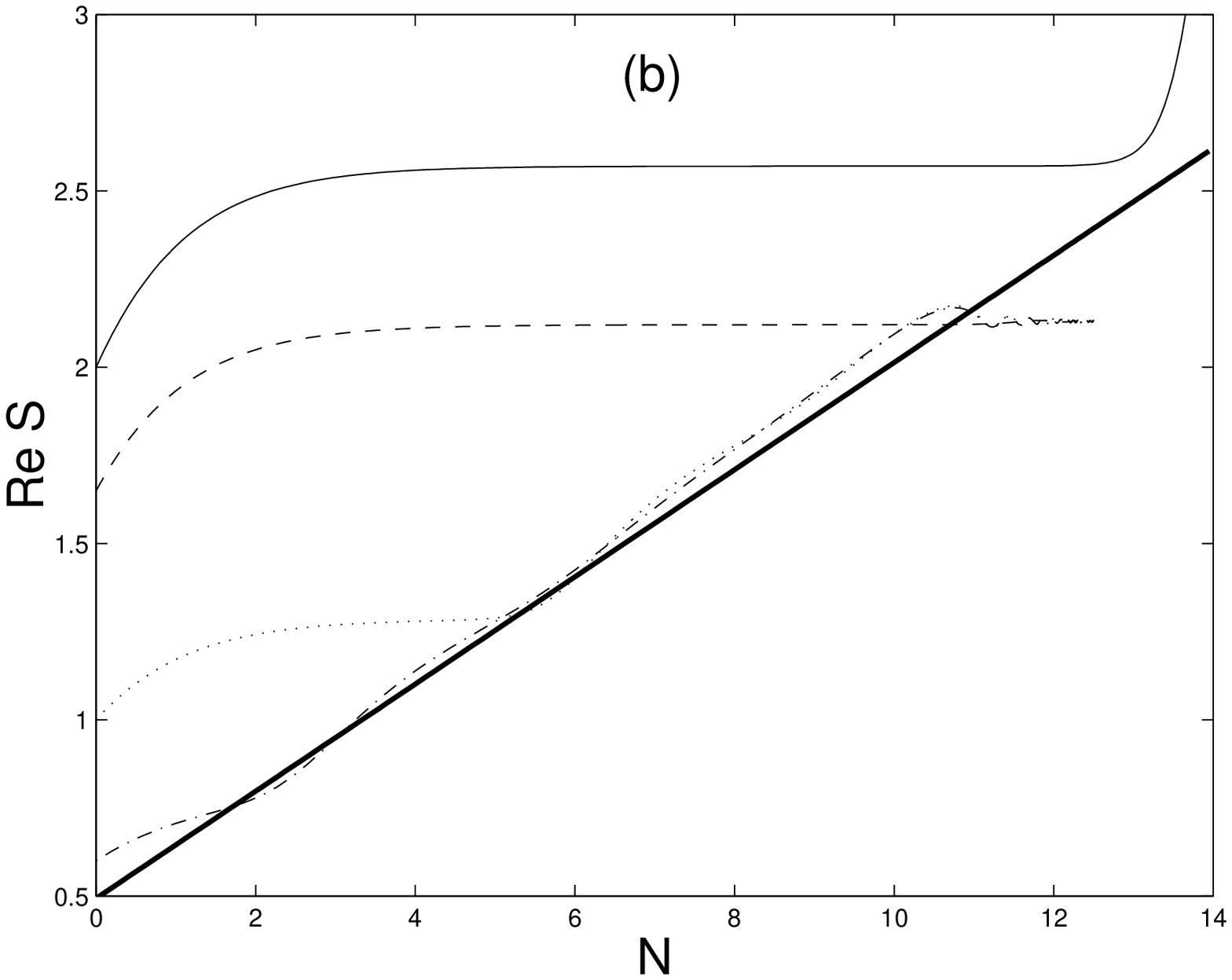,height=9cm,width=9cm,bbllx=0cm,bblly=7cm,bburx=21cm
,bbury=21cm}
}
\caption{}
{\footnotesize {\bf (a)} Solid line: plot of the scalar potential $V$ 
(in logarithmic units) vs ${\rm Re} S$ for two condensates with gauge 
groups SU(6) $\times$ SU(7) with $M_1=2$ and $M_2=8$ matter
representations repectively; dashed line: our exponential
approximation given by Eq.~(\ref{potap}). {\bf (b)} Evolution of 
${\rm Re}$ S vs $N$ for the two condensate potential plotted in (a) 
with $H_0=1$, $\gamma=4/3$, the different 
initial positions being ${\rm Re}S_{0} = 0.6$ (dot-dashes), 
${\rm Re}S_{0} = 1$ (dots), ${\rm Re}S_{0} = 1.65$ (dashes), 
${\rm Re}S_{0} = 2$ (solid) and the initial velocities given by 
${\rm Re}\dot{S}_{0} = {\rm Re}S_{0}/4$. The thick solid line 
represents the solution for the scaling regime Eq.~(\ref{stageIIe}), 
for the exponential potential of Eq.~(\ref{potap}).}
\end{figure}

We would like to study the evolution of the dilaton field (more
precisely of its real part) before it settles to the minimum. 
We have already seen that a single exponential scalar potential 
can be solved analytically leading to scaling solutions. However there
are major differences between this and the more realistic gaugino 
condensate models. First, the dilaton couples not only gravitationally
but also directly to the matter fields. For simplicity, and to avoid 
making any assumptions about specific models, we will neglect this 
effect throughout this paper. Second, in the case of two condensates, 
the superpotential Eq.~(\ref{sup}) contains two different exponentials
of ${\rm Re} S$, therefore the scalar potential Eq.~(\ref{pot}) will 
have all the different terms coming from $|W|^2$. In particular there 
will be a mixture of exponential and polynomial terms in the dilaton. 
And finally, the differential equation for $\phi$ written in 
Eq.~(\ref{system}) is meant to be obeyed by a canonically normalised 
field, and ${\rm Re} S$ is not. As it is well known, in the SUGRA 
Lagrangian the kinetic terms for scalar fields are given by 
$K_i^j D_{\mu} \phi_i D^{\mu} \bar{\phi}^j$, which in the case of 
${\rm Re} S$ introduces an extra factor of $1/(2 {\rm Re} S)^2$.
Therefore, the correct procedure in order to study the evolution of
the dilaton in an expanding Universe would be to solve the system
in terms of the canonically normalised field $\phi$, with ${\rm Re} 
S \equiv e^{\phi}$ or, alternatively, to modify Eq.~(\ref{system})
accordingly in order to account for the non canonical kinetic terms. 

Fortunately these problems can be overcome. First we note
that, even though the two condensates in Eq.~(\ref{sup}) have to be
carefully fine-tuned to produce a minimum for ${\rm Re} S$ at the
right value, for most of the evolution towards such a minimum only
one of them (the one with a smaller $\alpha_i$ value) will dominate.
For simplicity, we also keep the matter fields constant at their 
minimum value $A_{i \, {\rm min}} = (3 N_i /d_i  M_i)^{-2/\beta_i} 
\exp(-2\alpha_i S_{\rm min} /\beta_i) / \eta(T)^2$ during the 
evolution of ${\rm Re} S$. Therefore in the region we are studying, 
the superpotential can be approximated by, for example, the first 
condensate; moreover it is also easy to show that, in Eq.~(\ref{pot}),
the term proportional to $\partial W/\partial A_{1}$ dominates among 
those within the brackets. In conclusion, our scalar potential can be 
very well represented in the region before the minimum by the 
following expression
\be
V = \left| \frac{d_1 M_1 A_{1 {\rm min}}^{(M_1-N_1)/N_1}}{\sqrt{6} N_1 
(2 {\rm Re} T) \eta(T)^{\beta_1}} \right|^{2} e^{-2 \alpha_1 
{\rm Re} S} \; \;, 
\label{potap}
\ee
which corresponds to considering only the dominant term mentioned
above and setting ${\rm Re} S =1$ everywhere but in the exponential.
The result of such an approximation can be seen in Fig.~2(a),
represented by the dashed line (with, as everywhere in this article,
${\rm Re} T =1.2$), and it is good enough to justify our use of
Eq.~(\ref{potap}) when studying the evolution of ${\rm Re} S$ away 
from the minimum. 

Concerning the second problem, that of ${\rm Re} S$ not being a
canonically normalised field, we start by showing the exact 
numerical result of the evolution of the $\phi$ field,  
plotted in Fig.~2(b) in terms of ${\rm Re} S = e^{\phi}$ versus 
$N$, for a radiation dominated Universe (i.e. $\gamma=4/3$) and
initial conditions $H_0=1$ and ${\rm Re}\dot{S}_{0} = 
{\rm Re}S_{0}/4$. The different lines correspond to different initial 
conditions for ${\rm Re} S$. It is remarkable how the behaviour in the
first stages of the evolution is very similar to that shown in Fig.~1 
for a pure exponential potential, the difference appearing after 
$N \sim 11$ e-foldings and due to the presence of a minimum in this 
case. Depending on the initial position of the field it may or may not
fall in the minimum, but what is clear is that if the field reaches 
the scaling regime the former will certainly happen, and that occurs,
as we can see, for a very wide range of values of ${\rm Re} S_0$, 
contradicting the general belief that only for a very narrow range of 
initial values, all around the minimum, would the dilaton settle at 
its minimum. In fact the top curve of Fig.~2(b), which is the only one
that does not end up at the minimum, corresponds to an asymptotic 
value of the field ${\rm Re} \tilde{S}_0 \equiv e^{\tilde{\phi}_0}$ 
beyond the maximum of the potential.

We have checked that the scaling solution shown in Fig.~2(b) is very 
well represented by Eq.~(\ref{scaling}) in the case of the potential 
being given by Eq.~(\ref{potap}), i.e. $\lambda=2 \alpha_1$ and 
$V_0= | {d_1 M_1 A_{1 {\rm min}}^{(M_1-N_1)/N_1}}/{\sqrt{6} N_1 
(2 {\rm Re} T) \eta(T)^{\beta_1}} |^{2}$. That is, as if ${\rm Re} S$ 
were a canonically normalised field. This approximate solution 
corresponds to the thick solid line shown in Fig.~2(b).

We turn our attention to explaining this scaling behaviour of
the non canonical dilaton field by trying to solve Eq.~(\ref{system})
analytically for this case. In terms of the normalised dilaton,
$\phi$, the equations to solve are
\br
x'_{\phi} & = & -3x_{\phi} + \lambda e^{\phi} \sqrt{\frac{3}{2}} 
y_{\phi}^2 + \frac{3}{2} x_{\phi} [2x_{\phi}^2+ \gamma
(1-x_{\phi}^2-y_{\phi}^2) ] \nonumber \\
y'_{\phi} & = & -\lambda e^{\phi} \sqrt{\frac{3}{2}} x_{\phi} y_{\phi}
+  \frac{3}{2} y_{\phi} [2x_{\phi}^2+ \gamma (1-x_{\phi}^2-y_{\phi}^2)
] \label{systemfi} \\
H' & = & - \frac{3}{2} H [2x_{\phi}^2+ \gamma (1-x_{\phi}^2-y_{\phi}^2) 
] \;\;, \nonumber
\er
where $x_{\phi}$ and $y_{\phi}$ are the $x$ and $y$ of 
Eq.~(\ref{systemr}). Note that the presence of a more involved 
potential requires the replacement $\lambda \rightarrow \lambda 
e^{\phi}$. We can proceed as in the case of the pure exponential and 
solve for the two different stages defined before. Solving for Stage~I
is trivial, as the key point in this regime is to neglect any
dependence on the potential and therefore the solution (for $\phi$) is
identical to that in Eq.~(\ref{stageI}). Then
\be
{\rm Re} S_{\rm I} = {\rm Re} \tilde{S}_0 
\left(\frac{x_0}{\sqrt{1-x_0^2}} e^{-\frac32 (2-\gamma)N} + 
\sqrt{1+\frac{x_0^2}{1-x_0^2} 
e^{-3(2-\gamma)N}} \right)^{-\sqrt{\frac{8}{3}}/(2-\gamma)} \;\;,
\label{stageIs}
\ee
where ${\rm Re} \tilde{S}_0 \equiv e^{\tilde{\phi}_0}$, with 
$\tilde{\phi}_0$ given by Eq.~(\ref{phitil}). 
However solving for the scaling regime of Stage~II is much more 
complicated. To begin with, the form of the potential and the range of
values of $\phi$ we are interested in guarantees that $x, y \ll 1$, 
and therefore we can solve for $H$, obtaining the usual result 
$H=H_0 e^{-3\gamma N/2}$, once again indicating that the background 
fields are determining the evolution of the Universe. As for the other
two equations, let us rewrite them in terms of ${\rm Re} S$ and a new 
variable $x_S \equiv {\rm Re} S'/\sqrt{6}$ (note that
$y_S=y_{\phi}$). Again in the same approximation of small $x$, $y$ we 
have for the two first equations
\br
x'_S & = & - \frac{3}{2} (2-\gamma) x_S + \lambda \sqrt{\frac{3}{2}} 
({\rm Re} S)^2 y_S^2 \nonumber \\
y'_S & = & -\lambda \sqrt{\frac{3}{2}} x_S y_S + \frac{3}{2} \gamma
 y_S \;\;.
\label{systemS}
\er
Obviously, for the scaling regime observed in Fig.~2(b) (i.e. Stage
II) we must have $x'_{S}=y'_{S}=0$. The second equation in 
Eq.~(\ref{systemS}) will give us then the expected solution for $x_S$,
analogous to the pure exponential case
\be
x_S^c = \sqrt{\frac{3}{2}} \frac{\gamma}{\lambda} \;\;,
\ee
whereas from the first equation we find that
\be
y_S^2 = \frac{3 (2-\gamma) \gamma}{2 \lambda^2} 
\frac{1}{({\rm Re} S)^2} \;\;,
\label{ys}
\ee
that is, $y_S$ does not seem to reach a critical value but instead has
a dependence on $({\rm Re} S)^{-2}$. However, given the size of
$\lambda$ ($\geq 20$) and the range of values we are considering for
${\rm Re} S$ (between 0.3 and 2), the deviation of $y_S$ from its
expected critical value, given by $y_S^c = 3(2-\gamma)\gamma/(2 
\lambda^2)$, is not going to be significant, as it is obvious from
Fig.~2(b).
In any case let us compute this correction which 
modifies the pure scaling result by a factor of $\epsilon(N)$. Our
ansatz is then
\be
{\rm Re} S_{\rm II} = \frac{3\gamma}{\lambda} N + \frac{1}{\lambda} \ln 
\left( \frac{2V_0 \lambda^2}{9 H_0^2(2-\gamma)\gamma} \right) + 
\epsilon(N) \;\;.
\label{stageIIe}
\ee
Substituting into Eq.~(\ref{ys}) and using the definition of $y_S$
($\equiv \sqrt{V_0} e^{-\frac{\lambda}{2} {\rm Re} S}/(\sqrt{3} H)$) 
and the solution for $H$ we obtain
\be 
\epsilon(N) = -\frac{2}{\lambda} \ln \left[ \frac{3\gamma}{\lambda} 
N + \frac{1}{\lambda} 
\ln \left( \frac{2V_0 \lambda^2}{9 H_0^2(2-\gamma)\gamma} \right)
\right] \;\;,
\ee
which is indeed a very small numerical correction to the standard
result (in fact $x'_S = \epsilon''(N) \sim 0$), and the only
noticeable effect of having solved for a scalar field with an
exponential potential but non minimal kinetic terms, a very 
encouraging result.

The pure exponential approximation will eventually break down as the 
dilaton approaches its minimum, ${\rm Re}S_{\rm min}$. As can be seen 
from Fig.~2(b), if the field is in its scaling regime, it will not 
have enough energy to go over the maximum of the potential. Instead, 
it will oscillate around the minimum with an exponentially damped 
amplitude, settling down quickly to its final value. A simple 
estimation of the number of e-foldings needed to reach the minimum 
can then be obtained from the scaling solution Eq.~(\ref{stageIIe}) 
by equating ${\rm Re}S_{\rm II}$ to ${\rm Re}S_{\rm min}$. One obtains,
\be
N_{{\rm min}} = \frac{1}{3\gamma} \left[ \lambda {\rm Re} S_{\rm min} -
\ln \left( \frac{2V_0 \lambda^2}{9 H_0^2(2-\gamma)\gamma} \right)
\right] \;\;,
\label{nm}
\ee
(where we have ignored the $\epsilon$ correction as this result is 
accurate enough). Therefore once we have defined the example we are
working with and the background, we will have $N_{{\rm min}}$. In
fact, with this result we can also calculate $H_{{\rm min}}$
\be
H_{{\rm min}} = H_0 e^{-\frac{3}{2} \gamma N_{{\rm min}}} = \sqrt{ 
\frac{2V_0 \lambda^2}{9(2-\gamma)\gamma}} e^{-\frac{\lambda}{2} 
{\rm Re} S_{\rm min}}
\label{hm}
\ee
which, remarkably enough, is independent of $H_0$.

We have estimated $N_{{\rm min}}$ and $H_{{\rm min}}$ in a radiation 
dominated Universe ($\gamma=4/3$) for a number of hidden sector gauge 
groups with ${\rm Re} S \sim 2$ and $m_{3/2} \sim 1$ TeV, obtaining 
the almost invariant result
\br
N_{{\rm min}} & \sim & 11 \nonumber \\
H_{{\rm min}} & \sim & 5.10^{-10} M_P \;\;.
\er
Assuming a radiation dominated Universe, where $H \propto T^2/M_P$,
this implies $T_{{\rm min}} \sim 10^{13}$ GeV. The invariance of this 
result can be more or less understood by rewriting Eq.~(\ref{nm}) in 
terms of the gravitino mass and the gauge group factors of the two
condensates. The requirement of the model to be phenomenologically 
viable fixes most of those parameters leading to a constant value for 
$N_{{\rm min}}$ and, therefore, $H_{{\rm min}}$ and $T_{{\rm min}}$.

From these results it is possible to estimate the range of initial 
conditions that will eventually lead to a stable dilaton. As long as 
the field enters the scaling solution before reaching its minimum, we 
can ensure that it will be stabilized. For this to happen, a suficient 
condition is to take an initial value ${\rm Re}S_{0}$ between the 
scaling solution at $N=0$ (i.e. the constant term in 
Eq.~(\ref{stageIIe})) and ${\rm Re}S_{\rm min}$, and an initial
velocity such that the asymptotic solution ${\rm Re} \tilde{S}_0$ 
in  Eq.~(\ref{stageIs}), is smaller than ${\rm Re}S_{\rm min}$. 
Namely we obtain the following bounds for ${\rm Re}S_{0}$ and $x_{0}$,
\br
& & \frac{1}{\lambda} \ln 
\left( \frac{2V_0 \lambda^2}{9 H_0^2(2-\gamma)\gamma} \right)
 <   {\rm Re}S_{0}   < {\rm Re}S_{\rm min} \nonumber \\
& &  \label{inibound} \\ 
& & 0  <  x_{0}  <  \tanh\left[ \frac{\sqrt{6}}{4}(2-\gamma) 
\ln\left( \frac{{\rm Re}S_{\rm min}}{{\rm Re}S_{0}} \right) 
\right] \;\;. 
\nonumber
\er
Some examples can be seen in Fig.~2(b). Note that this region gets 
smaller for decreasing $H_{0}$, as expected. There will also be a few 
initial conditions outside these limits that will still lead to a 
stable dilaton, such as initial values between the maximum and the 
minimum, values close to the lower bound on ${\rm Re}S_{0}$ or 
allowing for negative initial velocities. Still, even if we do not 
take these into account, it is clear that there is a sizable region 
in parameter space that allows the dilaton to evolve to its minimum 
and stay there.

A further possibility is to consider an inflationary scenario. This 
will correspond to the case of a very small (and changing) $\gamma$. Note 
that it is the barotropic fluid that is producing inflation, so that
we are not considering any kind of dilaton driven inflation. We can
see from Eq.~(\ref{stageIIe}) that for $\gamma \simeq 0$ the scaling 
solution is practically horizontal. It would take a large amount of 
e-foldings for this solution to reach the minimum of the 
potential. Furthermore, for some models of inflation, $\gamma$ would be 
rapidly changing and one does not expect the field to follow the 
scaling solution exactly in these cases \cite{copel98}. Nevertheless, 
a few e-foldings of inflation can open up a large region of parameter 
space. Unless the energy density is completely dominated by the 
dilaton, the almost constant $H$ of an inflationary scenario will quickly
freeze the field at a constant value that will then lead to a scaling 
solution during reheating ($\gamma = 1$). The bound for $x_0$ in 
Eq.~(\ref{inibound}) still applies, but is maximized for $\gamma = 0$.
One then expects most of the region of parameter space to evolve to a 
stabilized dilaton if there is an initial small period of inflation.

\subsection{One condensate with Knp}

In this section we will perform the same analysis of the cosmological 
evolution of the dilaton field for a different class of models
proposed more recently \cite{casas96,binet97,barre97}. These consist
of a single condensate which gets stabilized by the presence of 
nonperturbative corrections to the K\"ahler potential, which form
has been suggested in \cite{shenk90}. Therefore we are dealing
now with a scalar potential given again by Eq.~(\ref{pot}), where
the superpotential is simply
\be
W = \frac{C}{\eta(T)^6} e^{-\alpha {\rm Re} S},
\ee
with $\alpha=8 \pi^2/N$ for a SU(N) group and $C=-N/(32 \pi^2 e)$. The
K\"ahler potential is now more involved, $K=K_0+K_{np}$, where $K_0$
is defined by Eq.~(\ref{kahler}) (with an overall modulus) and the 
nonperturbative correction is parameterised as
\be
K_{np} = \frac{D}{B\sqrt{{\rm Re} S}} 
\log\left( 1+e^{-B (\sqrt{{\rm Re} S}-\sqrt{S_0})} \right) 
\;\;,
\label{ours}
\ee
that is in terms of three constants $S_0$, $D$, and $B$, the first of
which just determines the value of ${\rm Re} S$ at the minimum. 
Therefore this description is effectively made in terms of only $D$
and $B$, which are positive numbers. It has been shown \cite{barre97} 
that for a wide range of values of both parameters it is possible to 
generate a minimum at $S_0$ with zero cosmological constant. The shape
of these potentials is again very similar to that of the two
condensate models.

Another feature of this ansazt for the nonperturbative corrections
is that it is very well approximated in the region ${\rm Re} S<S_0$
by the following expression:
\be
e^{K_{np}} =  e^{-D (\sqrt{{\rm Re} S}-\sqrt{S_0})/\sqrt{{\rm Re} S}} 
\;,
\label{Knp}
\ee
and therefore the scalar potential is given by
\be
V = \frac{e^{-D (1-\sqrt{S_0/{\rm Re} S})}}{2 {\rm Re}S} 
   \left[ \frac{4 (1- D \sqrt{S_0/{\rm Re} S} +2 \alpha {\rm Re} S)^2}
               {4 + 3D \sqrt{S_0/{\rm Re} S}} 
           -3  \right] 
   |W|^2 \;\;.
\label{potnp}
\ee
In this second example, even though we have a single condensate to
start with, the dependence of the potential upon ${\rm Re} S$ is
much more complicated, as can be expected from the form of the
nonperturbative corrections to $K$, given by Eq.~(\ref{Knp}). To be 
able to write Eq.~(\ref{potnp}) in the usual exponential form 
$V=V_0 e^{-\lambda {\rm Re} S}$ we set ${\rm Re} S = 1$ everywhere
except in the exponents for which we use a linear fit. We obtain
\be 
V \simeq e^D \frac{2 (D \sqrt{S_0} + 2 \alpha)^2}{(4+3
D \sqrt{S_0})} \left( \frac{C}{\eta(T)^6} \right)^2 e^{-(2 \alpha
+D/S_0) {\rm Re} S} \;\;,
\label{apnp}
\ee
that is a pure exponential, where the exponent depends on the value
of $D$. As it was mentioned before, for a given hidden gauge group, 
there exist a series of values of ($D$, $B$) for which $V$ has a
minimum at $S_0$ with zero cosmological constant. That means that the
cosmological evolution of a particular hidden sector interaction will
be different depending on which values of the pair ($D$, $B$) we are
considering. This can be clearly appreciated in Fig.~3 where we plot
the cosmological evolution of the dilaton field versus the number of
e-foldings $N$ for the same gauge group, SU(5), and initial conditions
$H_0=1$, $\gamma=1$, ${\rm Re}\dot{S_{0}} = 0.15$, ${\rm Re}S_0=0.6$.
\begin{figure}
\centerline{
\psfig{figure=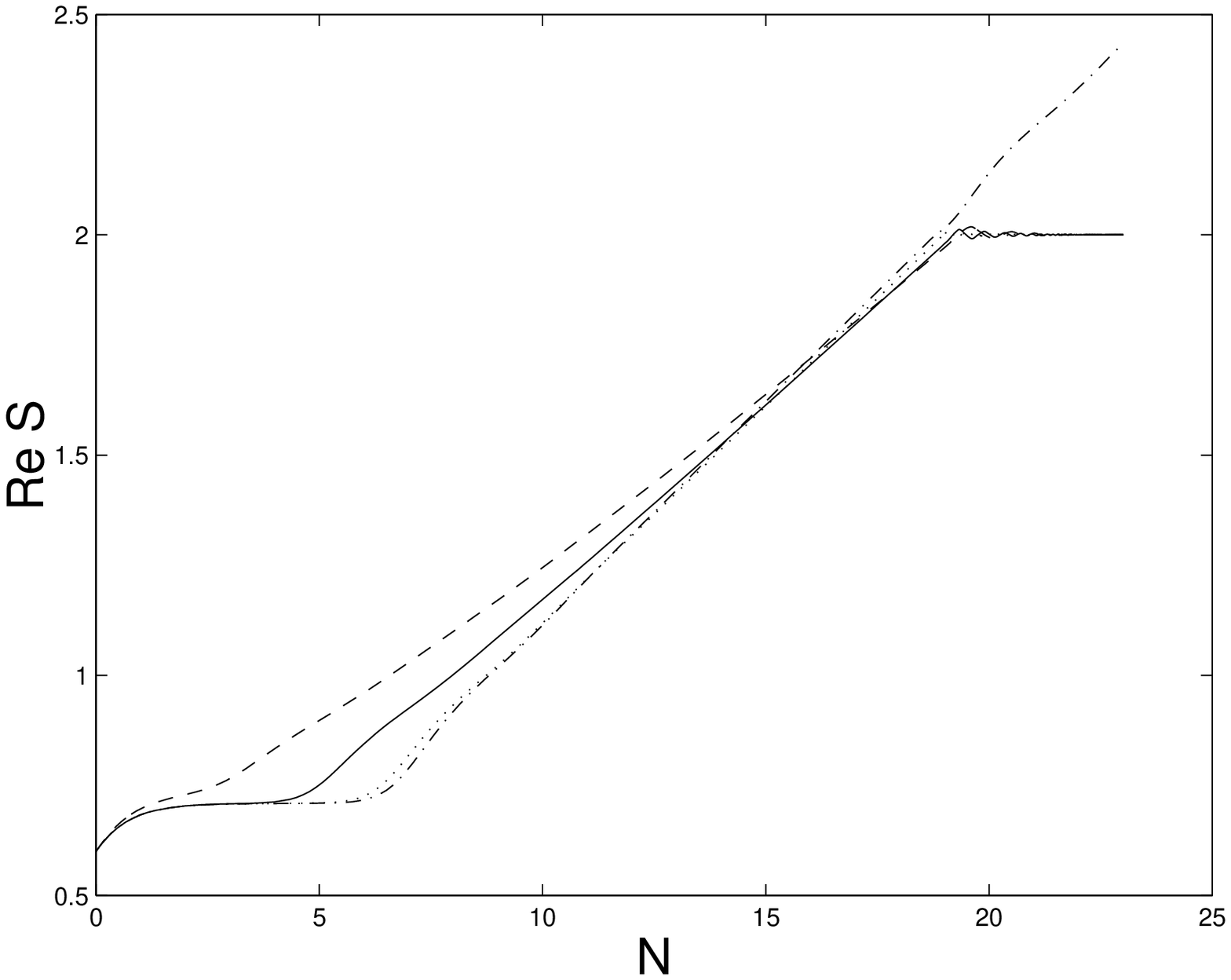,height=9cm,width=9cm,bbllx=0cm,bblly=7cm,bburx=21cm
,bbury=21cm}
}
\caption{}
{\footnotesize Evolution of ${\rm Re}$ S vs $N$ for the potential of
Eq.~(\ref{potnp}) with gauge group SU(5) and initial conditions $H_0=1$, 
$\gamma=1$, ${\rm Re}\dot{S_{0}} = 0.15$, ${\rm Re}S_0=0.6$. The
different lines correspond to different values for $D$ in Eq.~(\ref{ours}):
$D=1$ (dash-dotted), $D=3$ (dotted), $D=10$ (solid) and $D=20$
(dashed).}

\end{figure}
Depending on the value of $D$ (the corresponding $B$ is fixed in order
to have a zero cosmological constant at the minimum given by $S_0$) we
see that the field will fall into the minimum provided that this
minimum is not too fine-tuned. This is determined by the magnitude of 
$D$, as was described in \cite{barre97}, the smaller it is the bigger
the value of $B$ (and the amount of fine-tuning in the potential) is
to set $V=0$ at the minimum. Therefore in this particular case we see
how for $D \leq 1$ the minimum becomes too fine-tuned and is unable to
stop the field from rolling past the maximum. Unfortunately these 
small $D$ solutions are precisely the ones which correspond to a 
largest hierarchy between the dilaton and the gravitino masses 
\cite{barre97}, and therefore provide us with a solution to the 
``moduli problem'' for this field. However, for the range of values of
$D$ which give a satisfactory stabilization of the dilaton we can 
generate ratios between those two masses of up to $300$ which would, 
in most cases, be more than enough to solve the above-mentioned 
problem. Changing the value of $\gamma$ would correspond to a change 
in the minimal value of $D$ from which onwards the dilaton would fall 
in the minimum. That is, a bigger (smaller) value of $\gamma$ 
corresponds to a bigger (smaller) value of the minimal $D$, and 
therefore to a smaller (bigger) hierarchy between the dilaton and 
gravitino masses.
 
A similar analysis to the one performed in the previous section for
the two condensates model can be done here concerning the analytical 
solutions of the evolution equations (for which we use 
Eq.~(\ref{apnp})). Solving for Stage~I is identical as before (see 
Eq.~(\ref{stageIs})), as the system of differential equations given by 
Eq.~(\ref{systemr}) does not depend on the characteristics of the 
potential in this regime. For Stage~II we can obtain the analogous of 
Eq.~(\ref{stageIIe}) and in this case it is defined by 
$\lambda= 2 \alpha+D/S_0$ and $V_0 = e^D 2 (D \sqrt{S_0} + 2 \alpha)^2 
C^2/[(4 + 3 D \sqrt{S_0}) \eta^{12}]$. Expressions for the
number of e-foldings needed to reach the minimum, $N_{{\rm min}}$ as 
well as for  $H_{{\rm min}}$, are also obtained by replacing in 
Eqs.~(\ref{nm},\ref{hm}) the current expressions for $\alpha$ and 
$V_0$. Once again, the fact that we are imposing a consistent 
phenomenology (i.e., ${\rm Re} S \sim 2$ and $m_{3/2} \sim 1$ TeV) at 
low energies implies that the numerical values of $N_{{\rm min}}$
and $H_{{\rm min}}$ are very similar to those obtained in the previous 
section. In particular we can express $N_{{\rm min}}$ as
\be
N_{{\rm min}} = \frac{-2}{3 \gamma} \left[ \ln\left( \frac{m_{3/2}}{M_P}
\right) + \ln \left( \frac{(D \sqrt{S_0} + 2 \alpha) 2 \alpha}{
\sqrt{(4 + 3 D \sqrt{S_0})}} \right) - \ln \left(
\frac{3 H_0 \sqrt{(2-\gamma)\gamma}}{2\sqrt{2}} \right) \right] 
\;\;.
\ee
For the example shown in Fig.~3, namely $\gamma=1$, $H_0=1$, we
obtain, to a very good approximation, the almost invariant value 
$N_{{\rm min}} \sim 19$ which can be translated into 
$H_{{\rm min}} \sim 10^{-13}$.
 
\section{Conclusions}

We have examined the issue of the cosmological evolution of the
dilaton field in gaugino condensation models of SUSY breaking.
By studying the behaviour of a scalar field $\phi$ with an 
exponential potential, we have been able to obtain analytic 
expressions for its time evolution in the presence of the Hubble
parameter and a dominating background consisting of other matter 
fields. It turns out that under such circumstances the field $\phi$
tends to enter a scaling regime defined by constant values of both
the potential and the velocity of the field relative to the expansion
of the Universe, provided that this potential is steep enough.
 
Encouraged by these promising results we turned to apply them to
the dilaton field in two particular models of SUSY breaking: 
multiple gaugino condensation (or racetrack) and one condensate with 
nonperturbative corrections to the K\"ahler potential. However two 
major differences arise with respect to the ideal case we had dealt 
with before: the shape of these potentials is not exactly exponential 
and the dilaton field is not canonically normalised. We have shown
that these two problems can be easily overcome, and we have found 
very accurate analytic approximations to explain our numerical
results. In both cases it is possible, contrary to the general belief,
to stabilize the dilaton at its minimum for a large range of initial
conditions (i.e., initial values for its position and velocity with
$\gamma \sim 1$) despite of the steepness of the potentials in their 
strong coupling regime. A previous small period of inflation (i.e., 
$\gamma \sim 0$) opens up even more the region of parameter space 
leading to a stable dilaton. Moreover, the number of e-foldings 
required to do so and the value of the Hubble parameter at the minimum
seem to be fixed by the requirement of having a successful
phenomenology at low energies.

\section*{Acknowledgements}

The authors thank Andrew Liddle for interesting discussions. 
The work of T.~B. was supported by JNICT (Portugal). The work of 
B.~dC. and E.~C. was supported by PPARC.

\end{document}